\documentclass[
reprint,
amsmath,amssymb,
aps,
prb,
superscriptaddress,
floatfix,
]{revtex4-2}

\usepackage{graphicx}
\usepackage{bm}
\usepackage{epstopdf}
\usepackage{hyperref}
\hypersetup{colorlinks=true,linkcolor=blue,citecolor=blue,urlcolor=blue}

\begin{document}

\preprint{Preprint ID}

\title{Probing interacting two-level systems with rare-earth ions}

\author{Dapeng Ding}
\affiliation{Huygens-Kamerlingh Onnes Laboratory, Leiden University, 2333 CA Leiden, The Netherlands}
\author{David van Driel}
\affiliation{Huygens-Kamerlingh Onnes Laboratory, Leiden University, 2333 CA Leiden, The Netherlands}
\author{Lino M. C. Pereira}
\affiliation{KU Leuven, Instituut voor Kern- en Stralingsfysica, 3001 Leuven, Belgium}
\author{Jared F. Bauters}
\affiliation{Department of Electrical and Computer Engineering, University of California Santa Barbara, Santa Barbara, CA 93106, USA}
\author{Martijn J. R. Heck}
\affiliation{Department of Electrical and Computer Engineering, University of California Santa Barbara, Santa Barbara, CA 93106, USA}
\author{Gesa Welker}
\affiliation{Huygens-Kamerlingh Onnes Laboratory, Leiden University, 2333 CA Leiden, The Netherlands}
\author{Michiel J. A. de Dood}
\affiliation{Huygens-Kamerlingh Onnes Laboratory, Leiden University, 2333 CA Leiden, The Netherlands}
\author{Andr\'{e}~Vantomme}
\affiliation{KU Leuven, Instituut voor Kern- en Stralingsfysica, 3001 Leuven, Belgium}
\author{John E. Bowers}
\affiliation{Department of Electrical and Computer Engineering, University of California Santa Barbara, Santa Barbara, CA 93106, USA}
\author{Wolfgang L\"{o}ffler}
\affiliation{Huygens-Kamerlingh Onnes Laboratory, Leiden University, 2333 CA Leiden, The Netherlands}
\author{Dirk Bouwmeester}
\email{bouwmeester@physics.ucsb.edu}
\affiliation{Huygens-Kamerlingh Onnes Laboratory, Leiden University, 2333 CA Leiden, The Netherlands}
\affiliation{Department of Physics, University of California Santa Barbara, Santa Barbara, CA 93106, USA}

\date{\today}

\begin{abstract}

Two-level systems (TLS) in amorphous materials limit coherence times of a number of solid-state quantum devices. Interactions between TLS become prominent below 100 mK, but the coupling mechanism and statistical properties are still unclear. Here we determine the homogeneous linewidth of ytterbium ions (Yb\textsuperscript{3+}) in silica glass at 10--80 mK by using photon echo techniques as a probe of TLS. First, the homogeneous linewidth can be reduced by applying a magnetic field of 0.3 T. This effect is due to reduced magnetic interactions between adjacent Yb\textsuperscript{3+}. Secondly, we observe saturation of the linewidth below 50 mK to a level of approximately 30 kHz, which is much larger than the lifetime-limited value of 0.2 kHz. This saturation behavior is in conflict with the coupling to independent TLS. We show that this effect can be explained by coherently coupled TLS.
\begin{description}
\item[DOI]
\end{description}

\end{abstract}

\maketitle

\section{Introduction}

Two-level systems (TLS) or tunneling systems are single or groups of atoms in amorphous materials which can tunnel between two nearly degenerate configurations through a potential barrier. Individual TLS is characterized by the energy difference $ \epsilon $ and the coupling strength $ \Delta $ between the two configurations. The model of TLS describes independent TLS with a simple probability distribution of these two parameters and has successfully explained most of the low-temperature properties of amorphous materials since it was first conjectured in 1972 \cite{Anderson72,Phillips72} (referred to as the standard TLS model). Presence of TLS in quantum devices causes excess noise and therefore deteriorates their performance \cite{Muller19}. For example, TLS currently limit coherence times of superconducting quantum bits \cite{Muller15} and rare-earth-doped optical fibers as quantum memories \cite{Saglamyurek15a,Saglamyurek16}. Hence, in-depth understanding of TLS is key for addressing these issues.

Despite the success in general, the standard TLS model fails at temperatures below 100 mK. Significant discrepancies have been reported for a number of properties of amorphous materials such as sound velocity \cite{Vancleve94,Classen94}, internal friction \cite{Esquinazi92,Classen00,Konig02}, and dielectric constant \cite{Rogge96}. These experiments appear to indicate that TLS can no longer be treated independently as in the standard TLS model. Interaction between TLS is introduced to explain increased $ 1/f $ noise of a superconducting resonator \cite{Burnett14}. Single coupled TLS pairs have been observed in a Josephson junction \cite{Lisenfeld15}, but the coupling mechanism possibly due to either electric dipole interaction or strain-mediated elastic interaction is an open question and the probability distribution is still unclear.

Rare-earth-ion-doped crystals, glass, and optical fibers have long been used as laser media and optical amplifiers. The homogeneous linewidth of rare-earth ions are very sensitive to the properties of the host materials and thus excellent for probing TLS in amorphous materials. To date, experimental investigations on rare-earth ions below 100 mK are elusive. Hegarty \textit{et al.} first studied Nd\textsuperscript{3+}-doped fiber using two-pulse photon echo techniques at zero magnetic field \cite{Hegarty83}. They found that the homogeneous linewidth at 50 mK deviates from the trend at higher temperatures (0.1--1 K). Staudt \textit{et al.} observed a saturation behavior of the homogeneous linewidth of Er\textsuperscript{3+} below 100 mK using spectral hole burning techniques \cite{Staudt06}. However, the linewidth is strongly broadened by spectral diffusion and laser fluctuation as a limitation of the techniques. In these two cases, the homogeneous linewidth is broadened by multiple mechanisms and does not reflect the intrinsic properties of TLS.

In this article, we study the homogeneous linewidth $ \Gamma_\text{H} $ of trivalent rare-earth ions ytterbium (Yb\textsuperscript{3+}) in silica (SiO$ _2 $) glass as a probe of TLS using two-pulse photon echo techniques. We find that the linewidth can be reduced by applying a magnetic field of 0.3 T and the remaining linewidth saturates with decreasing temperature $ T $. The field-induced linewidth reduction is due to reduced magnetic interactions between adjacent Yb\textsuperscript{3+} and/or magnetic-dipole characters of TLS. We exclude the contribution from the interaction with the nuclear spins in the host materials and from the instantaneous spectral diffusion based on quantitative analysis. We are then able to unambiguously attribute the remaining linewidth to the intrinsic properties of TLS. We postulate that the saturation is due to the interaction between TLS which becomes prominent at low temperatures when phonon scattering is reduced. To this end, we calculate the density of states of coupled TLS pairs and show that this model leads to saturation of $ \Gamma_\text{H}(T) $. In addition, we show that the spectral diffusion is also reduced by the magnetic field using three-pulse photon echo techniques.

Our results are unique compared with previous studies \cite{Hegarty83,Staudt06,Macfarlane06,Sun06}, in that (i) the silica glass that is grown using low pressure chemical vapor deposition is of ultra-high purity and quality; (ii) most of the natural abundances of silicon (95.3\%) and oxygen (99.8\%) isotopes have zero nuclear spins; (iii) the isotope \textsuperscript{174}Yb with zero nuclear spin is selected in an implanter prior to the implantation into the glass; (iv) the peak doping concentration (10 ppm atom number) is two orders of magnitude lower than typical rare-earth-doped materials. These features indicate a benchmark study with well controlled impurities, nuclear spins, and mutual interactions, which provides insights into TLS beyond the standard TLS model.

The article is organized as follows. Sec.~\ref{sec:2pulse} describes the Yb\textsuperscript{3+}-doped device and the two-pulse photon echo experiment. Sec.~\ref{sec:linewidth} is devoted to the main results of this article, where we present the homogeneous linewidth of Yb\textsuperscript{3+} extracted from the two-pulse photon echo time traces as a function of temperature and magnetic field. Various line broadening mechanisms are discussed. In Sec.~\ref{sec:heat}, we verify the homogeneous linewidth at different experimental conditions in order to exclude heating effects induced by the excitation laser. We also study the heating effect through numerical simulation. In Sec.~\ref{sec:model}, we propose an explanation of the saturation of the homogeneous linewidth based on a model of coupled TLS. In Sec.~\ref{sec:3pulse}, we present the results of three-pulse photon echo and show that the spectral diffusion and energy relaxation rate of Yb\textsuperscript{3+} is also reduced by a magnetic field. In Sec.~\ref{sec:discussion}, limitations of this work and possible further studies are discussed. Finally, we summarize the article in Sec.~\ref{sec:conclusion}.

\section{Two-pulse photon echo}\label{sec:2pulse}

\begin{figure}[htb]
\centering
\includegraphics[width=1.0\columnwidth]{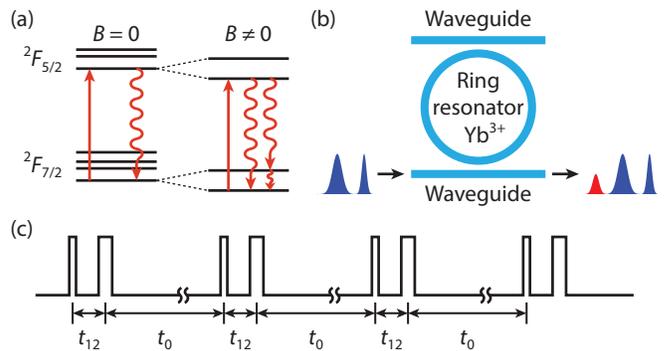}
\caption{(a) Energy level diagram (not to scale) of Yb\textsuperscript{3+} in silica glass with the relevant transitions indicated by arrows. For simplicity only the Zeeman splittings of the two lowest levels are drawn. (b) Schematic of an Yb\textsuperscript{3+}-doped ring resonator and measurement scheme. The ring resonator is coupled to two waveguides. Laser pulses are launched in one of the waveguides. Photon echo signals are measured through the same waveguide. (c) Laser pulse sequence for two-pulse photon echo measurements. Two pulses are separated by a delay time $ t_{12} $. An idle time $ t_0 $ that is comparable to energy relaxation times ensures sufficient population in the ground state for the next excitation cycle.}
\label{fig:device}
\end{figure}

The Yb\textsuperscript{3+}-doped silica glass is part of the cladding material of a silicon nitride (Si$ _3 $N$ _4 $) ring resonator. This device has been extensively studied for the Purcell effect \cite{Ding16}. Detailed description and characterization of the device, which will be only briefly summarized here, can be found therein. Yb ions are implanted into the silica cladding with an energy of 360 keV. The center of the Yb\textsuperscript{3+} spatial distribution (86 nm FWHM) is 72 nm to the interface between silica and Si$ _3 $N$ _4 $. This distribution avoids implantation damage and defects in Si$ _3 $N$ _4 $. The sample is then annealed at 950$ ^\circ $C for 3 h to remove implantation damage and optically activate Yb\textsuperscript{3+}. An energy level diagram of Yb\textsuperscript{3+} is shown in Fig.~\hyperref[fig:device]{\ref{fig:device}(a)}. We focus on the optical transitions centered at 976~nm which involve the lowest level of the ground state ($ ^2\text{F}_{7/2} $) and the lowest level of the excited state ($ ^2\text{F}_{5/2} $). These levels are doubly degenerate at zero magnetic field. Under a magnetic field, each level splits into two Zeeman levels. Figure~\hyperref[fig:device]{\ref{fig:device}(b)} illustrates a schematic of the device and the measurement scheme. Laser pulses at 976.0 nm are launched into a waveguide and subsequently coupled to the ring resonator. The ring resonator supports at least two modes with quality factors of $ 8.3\times10^5 $ and $ 4.8\times10^6 $, which will be referred to as low-$ Q $ and high-$ Q $ modes, respectively. Measurements are conducted by default through the low-$ Q $ mode unless explicitly specified. The Yb\textsuperscript{3+} in the ring resonator are excited and the fluorescence is coupled out through the same waveguide.

\begin{figure}[htb]
\centering
\includegraphics[width=1.0\columnwidth]{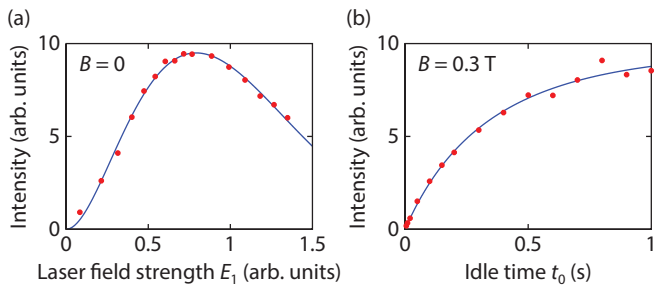}
\caption{(a) Two-pulse photon echo intensity at 10 mK and at zero magnetic field as a function of laser field strength showing an intensity maximum. The fit model is a sinusoidal function with an amplitude of exponential decay. (b) Two-pulse photon echo intensity at 10 mK and at $ B=0.3 $ T as a function of idle time $ t_0 $ showing a saturation behavior. The fit model is based on population analysis of a three-level system (see main text).}
\label{fig:echo_calib}
\end{figure}

The homogeneous linewidth of Yb\textsuperscript{3+} is measured by two-pulse photon echo techniques. The laser pulse sequence is displayed in Fig.~\hyperref[fig:device]{\ref{fig:device}(c)}. Two pulses with durations of $ t_1=60 $ ns and $ t_2=120 $ ns, respectively, are separated by a delay time $ t_{12} $. They are repeated with an idle time $ t_0 $ in between and photon echo signals are averaged. The peak power of the laser pulses is first calibrated at 10 mK and at zero magnetic field with $ t_{12}=0.5~\mu $s and $ t_0=5 $ ms. The measured data are shown in Fig.~\hyperref[fig:echo_calib]{\ref{fig:echo_calib}(a)}. A function $ I=I_1\exp(-\gamma E_1)\sin^2(\alpha E_1) $ is fitted to the data, where $ I_1 $ is a maximum intensity, $ \gamma $ is a decay rate, $ E_1 $ is the peak electric field strength of the laser pulses, and $ \alpha=dt_1/2\hbar $ with $ d $ being the transition dipole moment and $ \hbar $ being the reduced Planck constant. The exponential decay is due to inhomogeneous dephasing of the ions with a spectral distribution given by the linewidth of the laser pulses. A maximum intensity is reached when the two consecutive laser pulses perform $ \pi/2 $ and $ \pi $ operations for the ions, respectively. The corresponding laser power setting is used in the rest of this article.

At $ B=0.3 $ T, the Zeeman splittings of the ground and excited states are much larger than the laser linewidth. Because the upper Zeeman level of the ground state typically has much longer lifetime than the excited state \cite{Saglamyurek15b}, upon continuous strong excitation from the lower Zeeman level to the excited state, almost all the population will end up in the upper Zeeman level resulting in a depleted lower Zeeman level. Since the ions are repetitively excited, partial depletion occurs if $ t_0 $ is short compared to the lifetime $ T_1 $ of the upper Zeeman level.

The measured photon echo intensity at $ B=0.3 $ T as a function of $ t_0 $ is shown in Fig.~\hyperref[fig:echo_calib]{\ref{fig:echo_calib}(b)}. It increases with increasing $ t_0 $ and exhibits a saturation behavior. We model this process using population analysis of a three-level system and assume that an equilibrium is reached for a large number of pulses. The intensity is given by
\begin{equation}
I=\frac{1-\exp(-t_0/T_1)}{1-(1-\eta/2)\exp(-t_0/T_1)}I_2,
\end{equation}
where $ \eta $ is an excitation coefficient and $ I_2 $ is the saturation intensity. Because the measured echo intensity has not completely saturated, the fit yields large errors for the parameters. We are only able to determine the value of $ T_1 $ to be on the order of 1 s. In what follows, we use $ t_0=0.3 $ s at $ B=0.3 $ T as a tradeoff between measurement time and population in the lower Zeeman level ($ >50 $\%), while use $ t_0=5 $ ms at $ B=0 $ which is 6 times longer than the excited-state lifetime and ensures all the population has decayed to the ground state.

\section{Homogeneous linewidth and broadening mechanisms}\label{sec:linewidth}

\begin{figure}[thb]
\centering
\includegraphics[width=1.0\columnwidth]{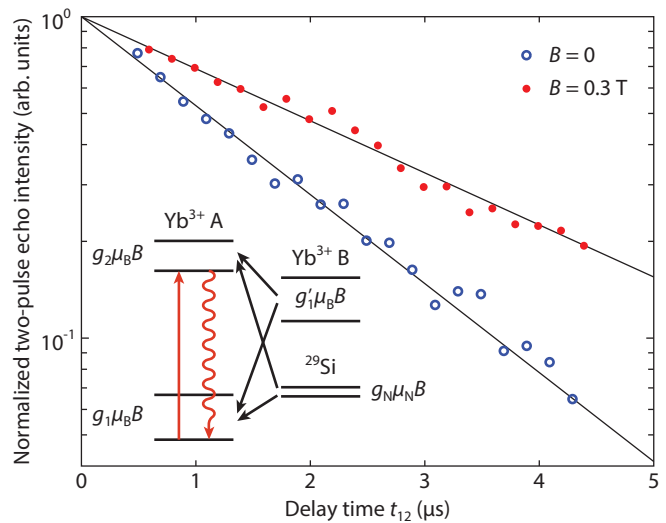}
\caption{Measured two-pulse photon echo intensities at 10 mK as a function of delay time $ t_{12} $ at $ B=0 $ (circles) and 0.3 T (dots) magnetic fields. The data are well described by exponential functions and are normalized by the values of the exponential functions at zero delay times. Inset: dipole-dipole interactions between two Yb\textsuperscript{3+} A and B, and between an Yb\textsuperscript{3+} A and a \textsuperscript{29}Si nucleus under an external magnetic field $ B $. These interactions can be viewed as a fluctuating magnetic field generated by Yb\textsuperscript{3+} B (ground state) or \textsuperscript{29}Si that modulates the ground and excited states of Yb\textsuperscript{3+} A. $ g_1 $ ($ g'_1 $) and $ g_2 $ are the $ g $ factors of the ground and excited states of Yb\textsuperscript{3+}, respectively. $ g_\text{N} $ is the $ g $ factor of \textsuperscript{29}Si nuclei and $ \mu_\text{B} $ ($ \mu_\text{N} $) is the Bohr (nuclear) magneton.}
\label{fig:echo_decay}
\end{figure}

For fixed $ t_0 $, the echo intensity decays with extended $ t_{12} $ due to homogeneous decoherence of the ions. The time constant of this decay is a measure of homogeneous linewidth $ \Gamma_\text{H} $. The measured echo intensities at 10 mK and at $ B=0 $ and 0.3 T as a function of $ t_{12} $ are shown in Fig.~\ref{fig:echo_decay}. They are well described by exponential functions in the form of $ I=I_3\exp(-4\pi \Gamma_\text{H} t_{12}) $ with $ I_3 $ and $ \Gamma_\text{H} $ being fitting parameters. The extracted values of $ \Gamma_\text{H} $ are $ 51\pm3 $ and $ 30\pm2 $ kHz at $ B=0 $ and 0.3 T, respectively.

The possible line broadening mechanisms that are responsive to a magnetic field include the magnetic dipole-dipole interactions between Yb\textsuperscript{3+} and between Yb\textsuperscript{3+} and nuclear spins of \textsuperscript{29}Si (+1/2, 4.7\% natural abundance) as depicted in the inset of Fig.~\ref{fig:echo_decay}. They are unrelated to TLS and similar effects are also observed in crystals \cite{Macfarlane97}. In a theory of electron paramagnetic resonance, the homogeneous linewidth $ \Gamma_{\text{H}A} $ of a magnetic dipole $ A $ due to the dipole-dipole interaction with a large number of magnetic dipoles $ B $ with random angle and position distributions can be written as \cite{Mims72}
\begin{equation}\label{eq:dipoleAB}
\Gamma_{\text{H}A}=(4\sqrt{3}/27)(\mu_0/\hbar) m_A m_B n_B,
\end{equation}
where $ \mu_0 $ is the vacuum permeability, $ m_A $ and $ m_B $ are the matrix elements of the magnetic moments of dipoles $ A $ and $ B $, respectively, and $ n_B $ is the concentration of dipole $ B $.

Adopting Eq.~\eqref{eq:dipoleAB} in our system, for the interaction between Yb\textsuperscript{3+}, the homogeneous linewidth reads
\begin{equation}
\Gamma_{\text{H}1}=(\sqrt{3}/54)(\mu_0/\hbar) |g_1-g_2| g_1 \mu_\text{B}^2 n_1,
\end{equation}
where $ g_1 $ and $ g_2 $ are the $ g $ factors of the ground and excited states of Yb\textsuperscript{3+}, respectively, $ \mu_\text{B} $ is the Bohr magneton, and $ n_1=3.3\times 10^{23} $ m$ ^{-3} $ is the concentration of Yb\textsuperscript{3+}. The values of $ g_1 $ and $ g_2 $ depend on the local atomic configurations of individual Yb\textsuperscript{3+} and in amorphous materials are expected to span the entire range of various crystal sites for their crystalline counterparts. However, the experimental data of $ g $ factors of Yb\textsuperscript{3+} are only available in crystalline CaF$ _2 $ as the host material \cite{Kirton69}. We use these data to make an order-of-magnitude estimation on the values of $ g_1 $ and $ g_2 $ to be 3.3 and 0.7, respectively, by averaging the values for seven different crystal symmetries along three directions. We obtain $ \Gamma_{\text{H}1}=95 $ kHz which is on the same order of magnitude as the measured homogeneous linewidth and field-induced linewidth reduction. The interaction between Yb\textsuperscript{3+} is almost completely suppressed under a magnetic field of 0.3 T because Yb\textsuperscript{3+} ions are frozen to their lower Zeeman levels with a large ratio of the Zeeman energy to the thermal energy ($ g_1\mu_\text{B}B/k_\text{B}T \approx 67 $, where $ k_\text{B} $ is the Boltzmann constant and $ T=10 $~mK).

Another model has been proposed for the field-induced linewidth reduction, in which TLS acquire a magnetic-dipole character through the coupling to rare-earth ions \cite{Macfarlane06,Boettger03}. A magnetic field creates an energy difference between the two potential wells. As a result, TLS is more favorable to remain in one well than the other such that the tunneling rate is reduced. In the present work, it is not possible to distinguish this model and the interaction between Yb\textsuperscript{3+}. Nevertheless, the thermal energy is much less than the Zeeman splittings of Yb\textsuperscript{3+} and therefore all the dephasing mechanisms related to magnetic dipoles of Yb\textsuperscript{3+} should in principle be suppressed. Moreover, we also observe a reduced energy relaxation and spectral diffusion rate by using three-pulse photon echo techniques. Those results are presented in Sec.~\ref{sec:3pulse}.

On the other hand, the suppression of the interaction between Yb\textsuperscript{3+} and nuclear spins of \textsuperscript{29}Si is much weaker because $ g_\text{N}\mu_\text{N}B/k_\text{B}T \approx 0.01 $, where $ g_\text{N} $ is the $ g $ factor of the nuclear spin of \textsuperscript{29}Si and $ \mu_\text{N} $ is the nuclear magneton. For this interaction, Eq.~\eqref{eq:dipoleAB} reads
\begin{equation}\label{eq:nuclear}
\Gamma_{\text{H}2}=(\sqrt{3}/54)(\mu_0/\hbar) |g_1-g_2| g_\text{N} \mu_\text{B}\mu_\text{N} n_2,
\end{equation}
where $ n_2=1.03\times 10^{27} $ m$ ^{-3} $ is the concentration of \textsuperscript{29}Si. We obtain that $ \Gamma_{\text{H}2}=54 $~kHz, which is on the same order of magnitude as $ \Gamma_{\text{H}1} $. The actual value of $ \Gamma_{\text{H}2} $ should be much smaller than that obtained by Eq.~\eqref{eq:nuclear} because the neighboring nuclear spins with a stronger influence on the Yb\textsuperscript{3+} flip slower than distant nuclear spins with a weaker influence (the frozen core effect \cite{Shelby78}). Here we would argue that this interaction is negligible in our system compared with other line broadening mechanisms based on the observed exponential decay of the echo time traces (Fig.~\ref{fig:echo_decay}). If the interaction with nuclear spins were significant compared to other mechanisms, the echo time trace would be non-exponential due to the frozen core effect. Previous experiments have shown that echo time traces merely due to the interaction with nuclear spins follow a universal expression of $ \exp\left[-(4\pi \Gamma_\text{H} t_{12})^{2.4}\right] $ for ruby and erbium ions in different crystals with different concentrations \cite{Ganem91}. In our experiment, the measured echo time traces as shown in Fig.~\ref{fig:echo_decay} which are well described by exponential functions exclude the interaction with nuclear spins as a major part in the value of $ \Gamma_\text{H} $.

\begin{figure}[thb]
\centering
\includegraphics[width=1.0\columnwidth]{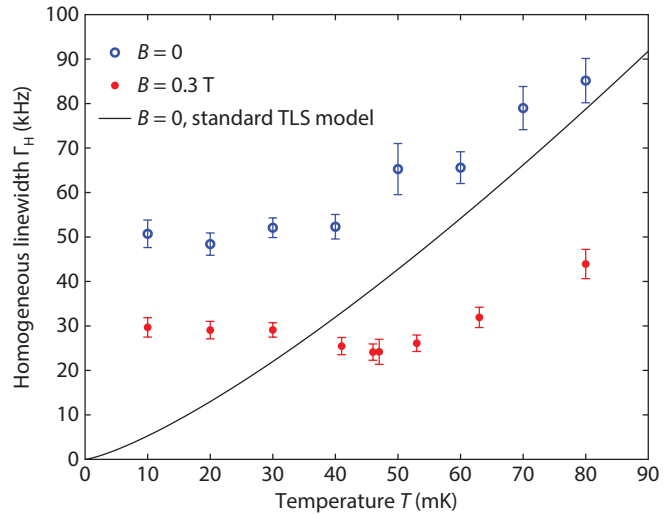}
\caption{Homogenous linewidth of Yb\textsuperscript{3+} extracted from the two-pulse photon echo measurements as a function of temperature at $ B=0 $ (circles) and 0.3 T (dots), respectively. The error bars represent 95\% confidence intervals (two standard deviations). The solid line follows the standard TLS model adopted from Ref.~\cite{Ding16}, using the homogeneous linewidth above 80 mK.}
\label{fig:linewidth}
\end{figure}

Instantaneous spectral diffusion (ISD) may also potentially cause additional line broadening \cite{Liu90}. The second laser pulse in the photon echo sequence excites Yb\textsuperscript{3+} and alters their magnetic moments through the different $ g $ factors of the ground and excited states. This process changes the magnetic interactions between Yb\textsuperscript{3+} after the second laser pulse and thus cannot be refocused by photon echo techniques. Based on Eq.~\eqref{eq:dipoleAB}, the homogeneous linewidth due to magnetic ISD is given by \cite{Liu90}
\begin{equation}\label{eq:ISD}
\Gamma_{\text{H}3}=(\sqrt{3}/108)(\mu_0/\hbar) (g_1-g_2)^2 \mu_\text{B}^2 n_3,
\end{equation}
where $ n_3 $ is the concentration of Yb\textsuperscript{3+} that are excited by the second laser pulse. Here $ n_3 $ is much smaller than total concentration $ n_1 $ due to an extremely large ratio of the inhomogeneous linewidth of Yb\textsuperscript{3+} $ \Gamma_\text{IH} $ (1.1 THz \cite{Ding16}) to the spectral bandwidth of the second laser pulse $ \Delta \nu_\text{L} $ (approximately 3.7 MHz, transform-limited) such that
\begin{equation}
n_3=n_1 \Delta \nu_\text{L}/\Gamma_\text{IH}=1.1\times 10^{18}~\text{m}^{-3}.
\end{equation}
We obtain that $ \Gamma_{\text{H}3}=0.13 $ Hz, a negligible contribution to the measured $ \Gamma_\text{H} $.

We now consider the effect of the electric dipoles of Yb\textsuperscript{3+}. The electric dipoles are highly polarized in the crystal field of the host material giving rise to the level splitting as shown in Fig.~\hyperref[fig:device]{\ref{fig:device}(a)} at $ B=0 $. Therefore the only possibility that the electric dipoles contribute to the line broadening is through excitation, in a similar way as the ISD caused by the magnetic dipoles. In analogy to Eqs.~\eqref{eq:dipoleAB} and \eqref{eq:ISD}, the homogeneous linewidth due to electric-dipole-induced ISD is given by
\begin{equation}\label{eq:electric}
\Gamma_{\text{H}4}=(\sqrt{3}/27)(1/(\varepsilon\hbar)) m_e^2 n_3,
\end{equation}
where $ \varepsilon=1.86\times 10^{-11} $ F/m is the absolute permittivity of silica and $ m_e $ is the difference of the permanent dipole moments between the ground and excited states. A typical value of $ m_e $ of rare earth ions is $ 6.6\times 10^{-31} \text{C}\cdot\text{m} $ (Stark shift 1 kHz/(V/m)) \cite{Macfarlane07}. We obtain that $ \Gamma_{\text{H}4}=16 $ Hz, which is a small value compared to the measured $ \Gamma_\text{H} $. Eventually, based on the above analysis, we conclude that the measured $ \Gamma_\text{H} $ is predominantly due to the interaction with TLS.

The values of $ \Gamma_\text{H} $ at various temperatures at $ B=0 $ and 0.3 T are summarized in Fig.~\ref{fig:linewidth}. The linewidth reduction of 20--40 kHz occurs in the entire range of 10--80 mK. The linewidth decreases with decreasing temperature and saturates below 50 mK. In Sec.~\ref{sec:heat}, we discuss possible heat sources that could bring the sample above the temperature of the cryostat and present the verification of the homogeneous linewidth for different experimental conditions. The fact that the saturation value is much larger than the lifetime-limited value of 0.2 kHz contradicts the standard TLS model which predicts a power law of $ T^{1.3} $ down to arbitrarily low temperatures \cite{Huber84,Broer86}, indicated by the solid black line in Fig.~\ref{fig:linewidth}. Weak temperature dependence of homogeneous linewidth has been reported for other rare-earth ions in glass at zero \cite{Hegarty83,Staudt06} and 1.3 T \cite{Staudt06}, but hitherto no explanation has been proposed. In Sec.~\ref{sec:model}, we show that this phenomenon could be explained by interactions between TLS.

\section{Laser heating}\label{sec:heat}

\begin{figure}[htb]
\centering
\includegraphics[width=1.0\columnwidth]{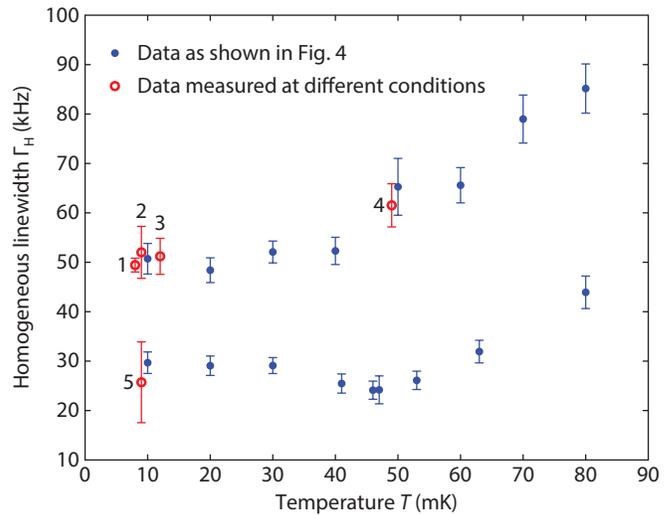}
\caption{Homogenous linewidth of Yb\textsuperscript{3+} extracted from the two-pulse photon echo measurements at different conditions (red with numbers) compared with the data shown in Fig.~\ref{fig:linewidth} (blue). The error bars represent 95\% confidence intervals (two standard deviations). The experimental conditions are listed in Table~\ref{tab:conditions}.}
\label{fig:linewidth2}
\end{figure}

The device is connected to high temperature components through long and thermally anchored optical fibers with an extremely low thermal conductivity. The device faces an upper plate of 50 mK with negligible radiation heating. The laser is the only possible heat source for the device. We verify the homogeneous linewidth at different experimental conditions in order to exclude heating effects induced by the laser. Five additional data points are shown in Fig.~\ref{fig:linewidth2} in addition to the data in Fig.~\ref{fig:linewidth}. Their experimental conditions are listed in Table~\ref{tab:conditions}. Despite different mean laser powers, the additional data are in good agreement with the data in Fig.~\ref{fig:linewidth}. The results indicate that heating effects on the device are negligibly small and support our explanation that the observed linewidth saturation is associated with TLS.

\begin{table}
\caption{Experimental conditions of the homogeneous linewidth data as shown in Fig.~\ref{fig:linewidth} and five additional data points as shown in Fig.~\ref{fig:linewidth2}. Peak power and mean power are the laser powers dissipated in the cryostat. These powers are estimated as the difference between the laser powers coupled in and out of the cryostat through optical fibers, only a fraction of which is absorbed in the sample.}
\begin{ruledtabular}
  \begin{tabular}{ c | c | c | c | c | c }
    Data & Field & Mode & Peak power & Idle time & Mean power\\ \hline
    Fig.~\ref{fig:linewidth} & 0 T & Low-$Q$ & 1.4 $\mu$W & 5 ms & 53 pW \\ \hline
    1 & 0 T & High-$Q$ & 2.7 $\mu$W & 5 ms & 96 pW \\ \hline
    2 & 0 T & Low-$Q$ & 1.4 $\mu$W & 10 ms & 26 pW \\ \hline
    3 & 0 T & Low-$Q$ & 1.4 $\mu$W & 10 ms & 26 pW \\ \hline
    4 & 0 T & Low-$Q$ & 1.4 $\mu$W & 10 ms & 26 pW \\ \hline
    Fig.~\ref{fig:linewidth}\footnote{An extra pulse of 10 $ \mu $s is added to each idle time for monitoring purposes.} & 0.3 T & High-$Q$ & 2.7 $\mu$W & 0.3 s & 96 pW \\ \hline
    5 & 0.3 T & Low-$Q$ & 1.5 $\mu$W & 0.3 s & 1 pW \\
  \end{tabular}
\end{ruledtabular}
\label{tab:conditions}
\end{table}

\begin{figure}[htb]
\centering
\includegraphics[width=1.0\columnwidth]{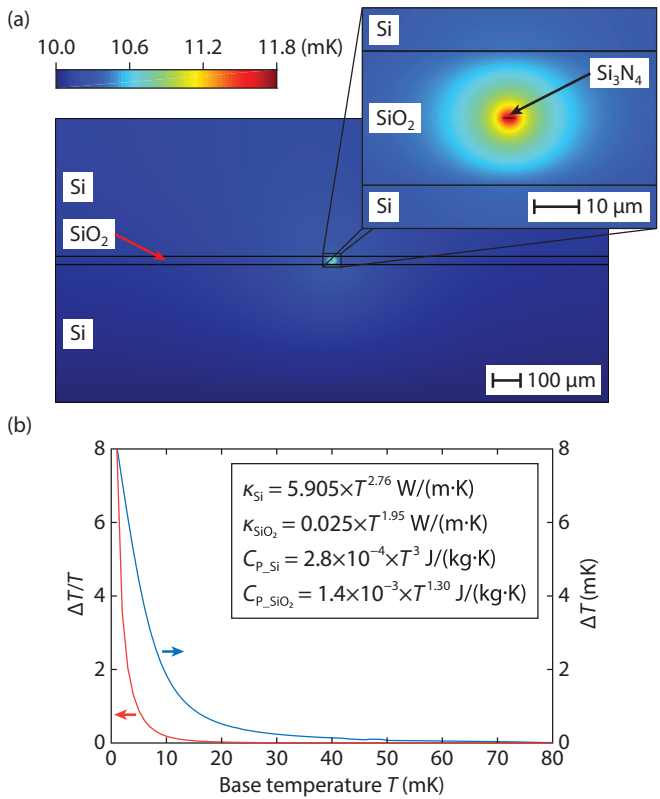}
\caption{Numerical simulation of the heating effect. (a) 2D area of a 3D axisymmetric model and simulated temperature distribution. The Si$ _3 $N$ _4 $ waveguide is modeled as a uniform heat source. The SiO$ _2 $ cladding is 30 $ \mu $m thick and is sandwiched in between two 0.5 mm-thick Si substrates. The temperature of the bottom of the substrate is a constant of 10 mK (base temperature). Inset: magnified area at the waveguide. (b) Temperature difference (blue curve) and relative difference (red curve) at various base temperatures. The relative difference drops to 2.6\% at 20 mK. Inset: temperature-dependent ($ T $ in Kelvin) thermal conductivity $ \kappa $ and specific heat $ C_\text{P} $ of Si and SiO$ _2 $ used in the simulation. These data are adopted from Refs.~\cite{Thompson61,Lasjaunias75,Keesom59} and assumed to be valid down to 1 mK.}
\label{fig:simulation}
\end{figure}

We simulate the heating effect by solving the steady-state heat equation using COMSOL. The model is 3D axisymmetric with the axis passing through the center of the ring resonator and normal to the substrate surface. Figure~\hyperref[fig:simulation]{\ref{fig:simulation}(a)} shows the 2D simulated area and the results. The Si$_3$N$_4$ waveguide with a width of 2.8 $\mu$m is modeled as a uniform heat source. We choose a heat power of 96 pW in the simulation that is the maximum laser power dissipated in the cryostat (not necessarily in the ring resonator) in all the experimental conditions in Table~\ref{tab:conditions}. The waveguide is embedded in the middle of 30 $ \mu $m-thick SiO$ _2 $ and sandwiched in between two 0.5 mm-thick Si substrates (wafer-bonding technique). The temperature-dependent thermal conductivity and specific heat data of Si and SiO$ _2 $ used in the simulation are given in the inset of Fig.~\hyperref[fig:simulation]{\ref{fig:simulation}(b)}. These data are adopted from Refs.~\cite{Thompson61,Lasjaunias75,Keesom59} and assumed to be valid down to 1 mK. The bottom of the substrate is kept at the base temperature of the cryostat, while the other boundaries are insulating.

Figure~\hyperref[fig:simulation]{\ref{fig:simulation}(a)} shows the simulation results at a base temperature of 10 mK. A temperature gradient centered at the waveguide is clearly visible. The temperature at the center of the waveguide is 11.8 mK, i.e., a difference ($ \Delta T $) of 1.8 mK and a relative difference ($ \Delta T/T $) of 0.18 compared to the base temperature. This value represents the upper bound of the heating effect at 10 mK. We conduct the simulation at various base temperatures. The temperature (relative) difference as a function of base temperature is shown in Fig.~\hyperref[fig:simulation]{\ref{fig:simulation}(b)}. The relative difference is substantial below 5 mK due to the extremely low thermal conductivity of Si and SiO$ _2 $ at this temperature scale. At 20 mK the relative difference drops to 2.6\%.

\section{Coupled TLS model}\label{sec:model}

In the standard TLS model, a TLS is described by the energy difference $ \epsilon $ between the two uncoupled potential wells and the coupling energy $ \Delta $ \cite{Anderson72,Phillips72}. The probability distribution function of independent TLS is
\begin{equation}\label{eq:P_single}
P(\epsilon,\Delta)=P_0/\Delta,
\end{equation}
where $ P_0 $ is a constant. The density of states is $ \rho(E) \propto E^{0.3} $, where $ E=\sqrt{\epsilon^2+\Delta^2} $ is the energy of the TLS. This density of states leads to $ \Gamma_\text{H}(T) \propto T^{1.3} $ by using $ E^\mu \rightarrow T^{1+\mu} $ which relates $ \rho(E) $ to $ \Gamma_\text{H}(T) $ \cite{Huber84}.

A. L. Burin and Yu. Kagan have theoretically studied coupling between two TLS through electrostatic interactions \cite{Burin94}. They demonstrate that coherently coupled TLS pairs form at sufficiently low temperatures when phonon scattering is weaker than the coupling strength within the pairs. Mathematically, a coupled TLS pair is equivalent to a single TLS with parameters $ \epsilon'=\epsilon_1-\epsilon_2 $ and $ \Delta'=(U_0/r^3_{12})\Delta_1\Delta_2/2\epsilon_1\epsilon_2 $, where $ \epsilon_1 $, $ \epsilon_2 $, $ \Delta_1 $, and $ \Delta_2 $ are the parameters of the two uncoupled TLS, $ r_{12} $ is the distance between the two TLS, and $ U_0 $ is a function of deformation potential, mass density, and sound velocity $ c $. TLS pairs have a drastically different probability distribution function compared with independent TLS:
\begin{equation}\label{eq:P_pair}
P'(\epsilon',\Delta')=\dfrac{P'_0 k_\text{B}T\delta(\epsilon')}{\Delta'^2}\Theta(\Delta'-U_0(k_\text{B}T/\hbar c)^3),
\end{equation}
where $ P'_0=\pi^3 P_0^2 U_0/12 $, $ \delta(x) $ is the Dirac delta function, and $ \Theta(x) $ is the Heaviside step function. Here we have added $ \delta(\epsilon') $ to the original form of Eq.~\eqref{eq:P_pair} in Ref.~\cite{Burin94} for resonant coupling between the two TLS, i.e., $ \epsilon'=0 $. The density of states of TLS pairs is calculated in the same way as the standard TLS model \cite{Phillips72}:
\begin{align}\label{eq:rho}
\rho'(E')&=\dfrac{\partial}{\partial E'} \left(\int_{-\infty}^{\infty} \text{d}\epsilon' \int_{\Delta'_\text{min}}^{\sqrt{E'^2-\epsilon'^2}} \text{d}\Delta'~P'(\epsilon',\Delta') \right)\nonumber\\
&=P'_0 k_\text{B}TE'^{-2},
\end{align}
where $ \Delta'_\text{min} $ is the minimum coupling strength that overcomes phonon scattering and $ E'=\sqrt{\epsilon'^2+\Delta'^2} $ is the energy of the pair. Equation~\eqref{eq:rho} leads to a temperature-independent $ \Gamma_\text{H} $, because temperature $ T $ cancels out according to $ E^\mu \rightarrow T^{1+\mu} $ with $ \mu=-2 $. Therefore this model leads to a saturation of $ \Gamma_\text{H}(T) $ from $ T^{1.3} $ to a constant with decreasing temperature, in qualitative agreement with our experimental observations.

\section{Three-pulse photon echo}\label{sec:3pulse}

\begin{figure}[htb]
\centering
\includegraphics[width=1.0\columnwidth]{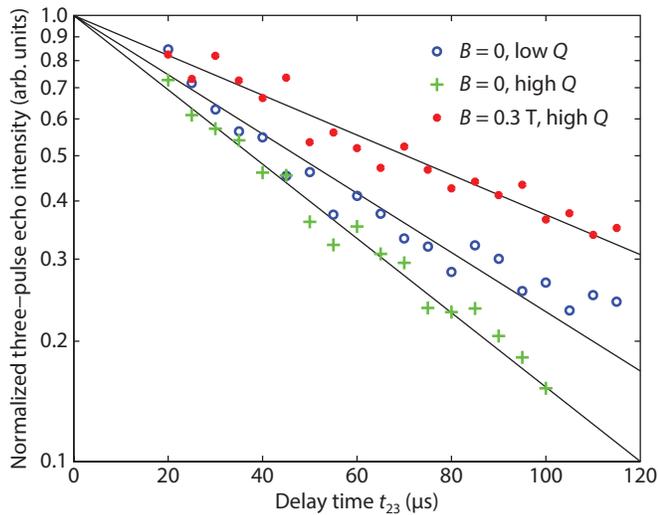}
\caption{Measured three-pulse photon echo intensities at 10 mK as a function of delay time $ t_{23} $ at $ B=0 $ (circles and plus signs) and 0.3 T (dots) magnetic fields for the laser on resonance with the low- (circles) and high-$ Q $ (plus signs and dots) modes of the ring resonator. The data are well described by exponential functions and are normalized by the values of the exponential functions at zero delay times.}
\label{fig:three_pulse}
\end{figure}

We conduct three-pulse photon echo experiments on the device. The three laser pulses are all 60 ns in width. The delay time between the first and second pulses $ t_{12} $ is a constant of 0.6 $ \mu $s, while the delay time between the second and third pulses $ t_{23} $ is varied. The first two pulses create a frequency comb in the absorption spectrum of Yb\textsuperscript{3+}. The third pulse is scattered by the frequency comb resulting in a photon echo with the same delay time as $ t_{12} $ after the third pulse. The lifetime of the frequency comb is sensitive to spectral diffusion and energy relaxation, but is insensitive to pure dephasing. Therefore it provides a method of studying other decoherence mechanisms than pure dephasing.

The measured three-pulse photon echo intensities as a function of $ t_{23} $ at 10 mK for the low- and high-$ Q $ modes of the ring resonator at $ B=0 $ and 0.3 T magnetic fields are shown in Fig.~\ref{fig:three_pulse}. They are well described by single-exponential functions in the form of $ I=I_4\exp(-\gamma_\text{SE} t_{23}) $, where $ \gamma_\text{SE} $ is a rate of spectral diffusion and energy relaxation. At $ B=0 $, the values of $ \gamma_\text{SE} $ are $ 15\pm2 $ and $ 18\pm1 $ ms$ ^{-1} $ for the laser on resonance with the low- and high-$ Q $ modes, respectively. The difference is due to different energy relaxation rates of the ions for different modes. At $ B=0.3 $ T, we obtain $ \gamma_\text{SE}=10\pm1 $ ms$ ^{-1} $ for the laser on resonance with the high-$ Q $ mode. This value is less than that at zero field by 8 ms$ ^{-1} $ due to reduced spectral diffusion. This effect is probably related to the magnetic-dipole properties of TLS as discussed in Sec.~\ref{sec:linewidth}. TLS have a broad distribution of tunneling timescales with the slow ones causing spectral diffusion. The spectral diffusion can be reduced by applying a magnetic field if TLS acquire a magnetic-dipole character \cite{Macfarlane06,Boettger03}.

\section{Discussion}\label{sec:discussion}

In this section, we shall discuss the limitations of this work and possible further studies. First of all, the Yb\textsuperscript{3+} ions are implanted close to the interface of the waveguide, potentially subject to strain and defects. The measured homogeneous linewidth may be different from the value in a bulk medium. Although this issue does not change the conclusion of this work about the role of TLS, caution should be exercised when comparing the homogeneous linewidth values here with those of bulk media.

Secondly, we choose an idle time of 0.3 s for all the homogeneous linewidth measurements at $ B=0.3 $ T, which is smaller than the lifetime of the upper Zeeman level of $ \sim $1 s. This implies that the Yb\textsuperscript{3+} with long lifetimes in their upper Zeeman levels are effectively excluded from the measurements. Therefore the measured homogeneous linewidth may be different from the averaged value of all the ions. Measurements of homogeneous linewidth at different idle times should be carried out in further studies.

In Sec.~\ref{sec:linewidth}, we quantitatively analyze different line broadening mechanisms in our experiment. These analyses can be verified by experiments. The argument that the magnetic dipole-dipole interaction between adjacent Yb\textsuperscript{3+} is frozen at $ B=0.3 $ T can be proven by measuring the homogeneous linewidth at various magnetic field strength. The homogeneous linewidth is expected to saturate at $ B=0.3 $ T. The negligibility of ISD can be verified by measuring the homogeneous linewidth at different laser frequencies over the inhomogeneous broadening of 1.1 THz. This method effectively varies the density of excited ions $ n_3 $ in Eqs.~\eqref{eq:ISD} and \eqref{eq:electric}. The homogeneous linewidth should not change if ISD is negligible.

In Sec.~\ref{sec:3pulse}, we show that the spectral diffusion and energy relaxation rate is reduced by a magnetic field of 0.3 T at 10 mK. It would be interesting to conduct the three-pulse photon echo experiment at various temperatures and to observe whether a similar saturation behavior appears as in the case of the homogeneous linewidth. The magnetic field dependence is also interesting as it may signify the relevant energy scale of the magnetic interaction.

\section{Conclusion}\label{sec:conclusion}

In summary, we investigate the homogeneous linewidth of Yb\textsuperscript{3+} in silica glass at ultra-low temperatures by using photon echo techniques. At zero magnetic field, the homogeneous linewidth saturates to a value of 50 kHz below 50 mK, which deviates from the prediction of the standard TLS model ($ T^{1.3} $). The saturation value contains contributions from the interaction with TLS and the interaction between Yb\textsuperscript{3+}. By applying a magnetic field of 0.3 T, the interaction between Yb\textsuperscript{3+} is suppressed and the saturation value is reduced to 30 kHz. Through quantitative analysis and additional measurements, we exclude the interaction with nuclear spins, instantaneous spectral diffusion, and heating effects as possible reasons for the saturation behavior and conclude that the interaction with TLS is the dominant line broadening mechanism. We show theoretically that the saturation behavior can be explained by emerging coherently coupled TLS pairs when phonon scattering is reduced below the coupling strength within the pairs. Coupling between TLS and the resulting density of states may have strong implications for quantum devices operating at ultra-low temperatures. In addition, we find that the rate of energy relaxation and spectral diffusion is also reduced by the magnetic field. This effect is probably related to the magnetic-dipole character of TLS. Further experimental investigations, e.g., temperature dependence, are needed to understand this effect.

\begin{acknowledgments}
This work is part of the research program financed by the Dutch Research Council (NWO). This work was supported by the NSF Quantum Foundry through Q-AMASE-i program Award No. DMR-1906325, the NWO Gravitation-grant Quantum Software Consortium - 024.003.037, DARPA MTO under the EPHI Contract No. HR0011-12-C-0006, Fund for Scientific Research-Flanders (FWO), and the KU Leuven (BOF-STRT/14/002).
\end{acknowledgments}

%

\end{document}